\documentclass[12pt]{article}
\usepackage{amsmath}
\usepackage{amssymb}

\numberwithin{equation}{section}
\newcommand{\pt}{$\cal PT$}
\newcommand{\ap}{a^{\dagger}}
\newcommand{\R}{\mathbb{R}}
\newcommand{\Z}{\mathbb{Z}}
\newcommand{\q}{{\cal Q}}
\DeclareMathOperator{\arctanh}{arctanh}
\newcommand{\bp}{b^{\dagger}}
\newcommand{\h}{{\cal H}}
\newcommand{\at}{\tilde{a}}
\newcommand{\atp}{\tilde{a}^{\dagger}}
\newcommand{\ds}{\displaystyle}
\newcommand{\xb}{\mbox{\boldmath $x$}}
\newcommand{\pb}{\mbox{\boldmath $p$}}
\newcommand{\sigmab}{\mbox{\boldmath $\sigma$}}
\newcommand{\llb}{\mbox{\boldmath $L$}}

\title{
Quasi-Hermitian supersymmetric extensions of a non-Hermitian oscillator Hamiltonian and of its 
generalizations}
\author{C Quesne\\ 
{\small Physique Nucl\'eaire Th\'eorique et Physique Math\'ematique,  Universit\'e Libre de Bruxelles,} \\ 
{\small Campus de la Plaine CP229, Boulevard~du Triomphe, B-1050 Brussels, Belgium}\\
{\small E-mail: cquesne@ulb.ac.be}}
\date{ }
\begin{document}
\baselineskip=22pt plus 1pt minus 1pt
\maketitle

\begin{abstract} 
A harmonic oscillator Hamiltonian augmented by a non-Hermitian \pt-symmetric part and its su(1,1) generalizations, for which a family of positive-definite metric operators was recently constructed, are re-examined in a supersymmetric context. Quasi-Hermitian supersymmetric extensions of such Hamiltonians are proposed by enlarging su(1,1) to a ${\rm su}(1,1/1) \sim {\rm osp}(2/2, \R)$ superalgebra. This allows the construction of new non-Hermitian Hamiltonians related by similarity to Hermitian ones. Some examples of them are reviewed.
\end{abstract}

\noindent
Short title: Supersymmetric extensions of a non-Hermitian oscillator

\noindent
Keywords: supersymmetry, non-Hermitian Hamiltonians, PT symmetry, superalgebras

\noindent
PACS Nos.: 03.65.Fd, 11.30.Pb
%
%
\newpage
\section{Introduction}

After the revival of the long-standing interest in non-Hermitian Hamiltonians, which followed the seminal paper of Bender and Boettcher \cite{bender} on a class of simple non-Hermitian $\cal PT$-symmetric Hamiltonians with a real and positive spectrum, many studies have been devoted to extending the framework of ${\cal N}=2$ supersymmetric quantum mechanics (SUSYQM) \cite{cooper, bagchi00}, the related intertwining operator method \cite{infeld} or the Darboux algorithm \cite{fatveev} to the non-Hermitian sector. SUSYQM has indeed provided a very useful technique for generating a lot of new exactly solvable (or quasi-exactly solvable) non-Hermitian (not necessarily \pt-symmetric) Hamiltonians with real (and/or complex) discrete eigenvalues by complexifying the underlying superpotential (see, {\sl e.g.}, \cite{cannata98, andrianov, znojil, cannata01, bagchi01, levai}). Such an approach has also been extended to some higher-order generalizations of SUSYQM (see, {\sl e.g.}, \cite{bagchi02, cannata03, bagchi05a, samsonov}).\par
%
%
Non-Hermitian Hamiltonians have been discussed in the general framework of pseudo-Hermiticity, which amounts to the existence in the relevant Hilbert space of a Hermitian invertible operator $\zeta$ such that $H^{\dagger} = \zeta H \zeta^{-1}$ \cite{mosta02a}. In such a context, the concept of pseudo-Hermitian SUSYQM has been introduced \cite{mosta02a, mosta02b, mosta04}\footnote{We prefer to use the full name `pseudo-Hermitian SUSYQM' instead of its often used abbreviated form `pseudo-SUSYQM' to avoid confusion with a different previous extension of SUSYQM designated by the latter name \cite{beckers, cq03}.} by replacing the superalgebra of standard SUSYQM by 
\begin{equation}
  \q^2 = \q^{\sharp 2} = 0, \qquad \left\{\q, \q^{\sharp}\right\} = 2 H_{\rm S},  \label{eq:pseudo-SUSY}   
\end{equation}
where all operators remain $\Z_2$-graded as usual, but $H_{\rm S}$ is pseudo-Hermitian with respect to some ($\Z_2$-graded) $\zeta_S$, while the supercharges $\q$ and $\q^{\sharp}$ are pseudo-adjoint of one another with respect to the same, {\sl i.e.}, $\q^{\sharp} = \zeta_S^{-1} \q^{\dagger} \zeta_S$. A generalization of this type of approach to higher order has also been proposed \cite{sinha, gonzalez}.\par
%
%
A consistent unitary theory of quantum mechanics for non-Hermitian Hamiltonians with a real spectrum can be formulated in terms of a positive-definite $\zeta$ (here denoted by $\zeta_+$) \cite{mosta02a, scholtz92, kretschmer}. Pseudo-Hermiticity is then termed quasi-Hermiticity: the Hamiltonian is Hermitian with respect to a new inner product of the Hilbert space, defined in terms of the metric operator $\zeta_+$, and it has a Hermitian counterpart $h = \rho H \rho^{-1}$ (where $\rho = \sqrt{\zeta_+}$) with respect to the original inner product. Similarly, in the SUSYQM approach, for a positive-definite $\zeta_{S+}$, equation (\ref{eq:pseudo-SUSY}) defines what we may call quasi-Hermitian SUSYQM, the supersymmetric Hamiltonian $H_{\rm S}$ having a ($\Z_2$-graded) Hermitian counterpart $h_{\rm S} = \rho_{\rm S} H_{\rm S} \rho_{\rm S}^{-1}$ (with $\rho_{\rm S} = \sqrt{\zeta_{S+}}$).\par
%
%
Many features of non-Hermitian Hamiltonians with a real spectrum have been studied \cite{swanson, geyer, jones, bagchi05b, scholtz06a, scholtz06b, musumbu} on a harmonic oscillator Hamiltonian augmented by a \pt-symmetric part, first proposed by Swanson \cite{swanson}. It is defined by
\begin{equation}
  H = \omega \bigl(\ap a + \tfrac{1}{2}\bigr) + \alpha a^2 + \beta a^{\dagger2},  \label{eq:swanson}
\end{equation}
where $\ap = (\omega x - {\rm i} p)/\sqrt{2\omega}$ and $a = \bigl(\ap\bigr)^{\dagger}$ are standard harmonic oscillator creation and annihilation operators (with $p = - {\rm i} d/dx$ and $\hbar = m = 1$), while $\omega$, $\alpha$, $\beta$ are three real parameters such that $\alpha \ne \beta$ and $\Omega^2 =\omega^2 - 4 \alpha \beta > 0$. This Hamiltonian is known to have a real, positive and discrete spectrum given by $E_n = \left(n + \frac{1}{2}\right) \Omega$, $n=0$, 1, 2,~\ldots.\par
%
%
{}For $H$ defined in (\ref{eq:swanson}), a family of positive-definite metric operators $\zeta_+(z)$, depending on a continuous variable $z \in [-1, 1]$, has been constructed \cite{musumbu} by a generalized Bogoliubov transformation approach. In addition, for each $\zeta_+(z)$ the corresponding equivalent Hermitian Hamiltonian $h(z)$ has been built and an additional observable $O(z)$, forming an irreducible set with $H$ and therefore fixing the choice of the metric \cite{scholtz92}, has been determined.\par
%
%
In a recent work \cite{cq07}, we have proposed an alternative derivation of these results, based on the fact that the Hamiltonian (\ref{eq:swanson}) can be written as a linear combination of su(1,1) generators $K_0$, $K_+$, $K_-$, characterized by the commutation relations
\begin{equation}
  [K_0, K_{\pm}] = \pm K_{\pm}, \qquad [K_+, K_-] = - 2 K_0,  \label{eq:com}
\end{equation}
and the Hermiticity properties
\begin{equation}
  K_0^{\dagger} = K_0, \qquad K_{\pm}^{\dagger} = K_{\mp}.  \label{eq:Hermite}
\end{equation}
One may indeed write
\begin{equation}
  H = 2\omega K_0 + 2\alpha K_- + 2\beta K_+  \label{eq:swanson-bis}
\end{equation}
with
\begin{equation}
  K_0 = \tfrac{1}{2} \bigl(\ap a + \tfrac{1}{2}\bigr), \qquad K_+ = \tfrac{1}{2} a^{\dagger2}, \qquad
  K_- = \tfrac{1}{2} a^2.  \label{eq:swanson-K}
\end{equation}
\par
%
%
Our derivation, being independent of the su(1,1) generator realization (\ref{eq:swanson-K}), has extended the results of \cite{musumbu} to all Hamiltonians that can be written in the form (\ref{eq:swanson-bis}) with $\omega, \alpha, \beta \in \R$, $\alpha \ne \beta$ and $\Omega^2 = \omega^2 - 4 \alpha \beta > 0$. So we have established that such Hamiltonians admit positive-definite metric operators $\zeta_+ = \rho^2$ with
\begin{equation}
  \rho = \exp\{\epsilon [2 K_0 + z (K_+ + K_-)]\},  \label{eq:rho-exp}
\end{equation}
where
\begin{equation*}
  \epsilon = \frac{1}{2\sqrt{1 - z^2}} \arctanh \frac{(\alpha - \beta) \sqrt{1 - z^2}}{\alpha + \beta -
  z \omega}, \qquad -1 \le z \le 1,
\end{equation*}
or, on other words,
\begin{equation}
  \rho = \left(\frac{\alpha + \beta - \omega z + (\alpha - \beta) \sqrt{1 - z^2}}{\alpha + \beta - \omega 
  z - (\alpha - \beta) \sqrt{1 - z^2}}\right)^{[2K_0 + z (K_+ + K_-)]/(4 \sqrt{1 - z^2})}.  \label{eq:rho}
\end{equation}
The corresponding equivalent Hermitian Hamiltonian and additional observable are given by
\begin{equation}
  h = \frac{1}{2\omega} \bigl[\nu (2K_0 + K_+ + K_-) + \mu \omega^2 (2K_0 - K_+ - K_-)\bigr] 
    \label{eq:h} 
\end{equation}
and
\begin{equation*}
  O = 2K_0 + z (K_+ + K_-), 
\end{equation*}
respectively. In (\ref{eq:h}), $\mu$ and $\nu$ are defined by
\begin{align*}
  \mu &= [(1+z) \omega]^{-1} \left[\omega - (\alpha + \beta) z - (\alpha + \beta - \omega z) \left(1 -
  \frac{(\alpha - \beta)^2 (1 - z^2)}{(\alpha + \beta - \omega z)^2}\right)^{1/2} \right], \\
  \nu &= (1-z)^{-1} \omega \left[\omega - (\alpha + \beta) z + (\alpha + \beta - \omega z) \left(1 -
  \frac{(\alpha - \beta)^2 (1 - z^2)}{(\alpha + \beta - \omega z)^2}\right)^{1/2} \right].
\end{align*}
\par
%
%
The aim of the present paper is to provide a quasi-Hermitian supersymmetric extension of the non-Hermitian oscillator Hamiltonian (\ref{eq:swanson}) and of its generalizations (\ref{eq:swanson-bis}) valid for any $\zeta_+(z)$. For such a purpose, we shall have to enlarge the su(1,1) Lie algebra considered in \cite{cq07} to an ${\rm su}(1,1/1) \sim {\rm osp}(2/2, \R)$ Lie superalgebra. This programme is carried out for the former Hamiltonian in section 2 and generalized to the latter one in section 3. Finally, section 4 contains the conclusion.\par
%
%
\section{Quasi-Hermitian supersymmetric extension of the non-Hermitian oscillator}

Since the Hamiltonian (\ref{eq:swanson}) has the same spectrum as a harmonic oscillator of frequency $\Omega$, it is clear that we may look for a quasi-Hermitian supersymmetric extension of the type
\begin{equation}
  H_{\rm S} = H_{\rm B} + H_{\rm F}, \qquad H_{\rm B} = H, \qquad H_{\rm F} = \Omega \left(\bp b - 
  \tfrac{1}{2}\right),  \label{eq:swanson-SUSY} 
\end{equation}
where $\bp$ and $b$ are fermionic creation and annihilation operators with $b^2 = b^{\dagger 2} = 0$, $\{b, \bp\} = 1$ and $[b, a] = [b, \ap] = [\bp, a] = [\bp, \ap] = 0$. This Hamiltonian has indeed the same spectrum $\Omega (n_{\rm B} + n_{\rm F})$, $n_{\rm B} = n = 0$, 1, 2,~\ldots, $n_{\rm F} = 0$, 1, as the boson-fermion harmonic oscillator \cite{bagchi00}. It acts on the ($\Z_2$-graded) boson-fermion Fock space $\h_{\rm S} = \h_{\rm B} \otimes \h_{\rm F}$. The metric operator, corresponding to (\ref{eq:rho}) and acting on  $\h_{\rm B}$, can be trivially extended to the whole space $\h_{\rm S}$ and we shall keep the same symbol to denote the extended operator. It is then obvious that the Hamiltonian (\ref{eq:swanson-SUSY}) is quasi-Hermitian with respect to the latter and that the equivalent Hermitian supersymmetric Hamiltonian is given by
\begin{equation*}
  h_{\rm S} = h_{\rm B} + h_{\rm F}, \qquad h_{\rm B} = h, \qquad h_{\rm F} = H_{\rm F}.
\end{equation*}
\par
%
%
Such a Hermitian Hamiltonian, with spectrum $\Omega (n_{\rm B} + n_{\rm F})$, can be explicitly written as a boson-fermion oscillator by expressing $h_{\rm B}$ as
\begin{equation*}
  h_{\rm B} = \Omega \left(\atp \at + \tfrac{1}{2}\right),
\end{equation*}
where
\begin{equation}
  \atp = \frac{1}{2\sqrt{\omega}} \biggl\{\biggl[\left(\frac{\nu}{\mu}\right)^{1/4} - \omega \left(\frac{\mu}  
  {\nu}\right)^{1/4}\biggr] a + \biggl[\left(\frac{\nu}{\mu}\right)^{1/4} + \omega \left(\frac{\mu}{\nu}
  \right)^{1/4}\biggr] \ap \biggr\}  \label{eq:a-tilde} 
\end{equation}
and $\at = \left(\atp\right)^{\dagger}$ satisfy bosonic commutation relations. As it is well known \cite{bagchi00}, $h_{\rm S}$ and the supercharge operators
\begin{equation}
  Q = \sqrt{2\Omega}\, \atp b, \qquad Q^{\dagger} = \sqrt{2\Omega}\, \at b^{\dagger}  \label{eq:Q}
\end{equation}
satisfy the standard SUSYQM superalgebra
\begin{equation*}
  Q^2 = Q^{\dagger2} = 0, \qquad \{Q, Q^{\dagger}\} = 2 h_{\rm S}.
\end{equation*}
\par
%
%
{}From the mutually adjoint supercharges (\ref{eq:Q}), we can now construct supercharges $\q$ and $\q^{\sharp}$ fulfilling equation (\ref{eq:pseudo-SUSY}) with the pseudo-Hermitian Hamiltonian (\ref{eq:swanson-SUSY}) and corresponding to a given $\zeta_+$ in (\ref{eq:rho}). It is indeed easy to check that such operators can be expressed as
\begin{equation}
  \q = \rho^{-1} Q \rho, \qquad \q^{\sharp} = \rho^{-1} Q^{\dagger} \rho.  \label{eq:pseudo-Q}
\end{equation}
\par
%
%
Then using equations (\ref{eq:a-tilde}), (\ref{eq:Q}) and the inverse of the generalized Bogoliubov transformation of \cite{musumbu}, 
\begin{align*}
 \rho^{-1} a \rho &= \left(\cosh\theta + \epsilon \frac{\sinh\theta}{\theta}\right) a + z \epsilon
    \frac{\sinh\theta}{\theta} \ap, \\
 \rho^{-1} a^{\dagger} \rho &= \left(\cosh\theta - \epsilon \frac{\sinh\theta}{\theta}\right) \ap - z \epsilon
    \frac{\sinh\theta}{\theta} a,     
\end{align*}
with $\theta = |\epsilon| \sqrt{1 - z^2}$, equation (\ref{eq:pseudo-Q}) acquires the form
\begin{equation}
  \q = \sigma W_+ + \tau W_-, \qquad \q^{\sharp} = \varphi V_- + \chi V_+,  \label{eq:pseudo-Q-bis}
\end{equation}
where $V_{\pm}$, $W_{\pm}$ are expressed in terms of the original bosonic and fermionic operators $\ap$, $a$, $\bp$, $b$ as
\begin{equation}
  V_+ = \frac{1}{\sqrt{2}} \ap \bp, \qquad V_- = \frac{1}{\sqrt{2}} a \bp, \qquad W_+ = \frac{1}{\sqrt{2}} 
  \ap b, \qquad W_- = \frac{1}{\sqrt{2}} a b.  \label{eq:V-W}
\end{equation}
In (\ref{eq:pseudo-Q-bis}), $\sigma$, $\tau$, $\varphi$ and $\chi$ are $z$-dependent real parameters, given by
\begin{equation}
\begin{split}
  \sigma &= \frac{1}{\sqrt{\omega}} \left\{(\omega \sqrt{\mu} + \sqrt{\nu}) \cosh\theta - [(\omega 
     \sqrt{\mu} + \sqrt{\nu}) + z (\omega \sqrt{\mu} - \sqrt{\nu})] \epsilon \frac{\sinh\theta}{\theta}\right\},
     \\
  \tau &= \frac{1}{\sqrt{\omega}} \left\{- (\omega \sqrt{\mu} - \sqrt{\nu}) \cosh\theta - [(\omega 
     \sqrt{\mu} - \sqrt{\nu}) + z (\omega \sqrt{\mu} + \sqrt{\nu})] \epsilon \frac{\sinh\theta}{\theta}\right\},
     \\
  \varphi &= \frac{1}{\sqrt{\omega}} \left\{(\omega \sqrt{\mu} + \sqrt{\nu}) \cosh\theta + [(\omega 
     \sqrt{\mu} + \sqrt{\nu}) + z (\omega \sqrt{\mu} - \sqrt{\nu})] \epsilon \frac{\sinh\theta}{\theta}\right\},
     \\
  \chi &= \frac{1}{\sqrt{\omega}} \left\{- (\omega \sqrt{\mu} - \sqrt{\nu}) \cosh\theta + [(\omega 
     \sqrt{\mu} - \sqrt{\nu}) + z (\omega \sqrt{\mu} + \sqrt{\nu})] \epsilon \frac{\sinh\theta}{\theta}\right\}.
\end{split}  \label{eq:sigma}
\end{equation}
\par
%
%
The operators $\q$ and $\q^{\sharp}$ of equation (\ref{eq:pseudo-Q-bis}) can alternatively be expressed in terms of $x$ and $p$ as
\begin{equation}
  \q = \frac{1}{2\sqrt{\omega}} [(\sigma + \tau) \omega x - {\rm i} (\sigma - \tau) p], \qquad
  \q^{\sharp} = \frac{1}{2\sqrt{\omega}} [(\varphi + \chi) \omega x + {\rm i} (\varphi - \chi) p].
  \label{eq:pseudo-Q-ter}
\end{equation}
\par
%
%
As illustrations, let us consider the three specific metric operators discussed in the literature \cite{geyer, jones, bagchi05b, scholtz06a, scholtz06b} and presented in \cite{musumbu}, corresponding to $z=0$, $z=1$ and $z=-1$, respectively. For each of them, we give below the values of the parameters (\ref{eq:sigma}) and the realization of the supercharges (\ref{eq:pseudo-Q-ter}), satisfying equation (\ref{eq:pseudo-SUSY}) with the pseudo-Hermitian Hamiltonian (\ref{eq:swanson-SUSY}):

\noindent (i) for $z=0$, $\epsilon = (1/4) \log(\alpha/\beta)$ and thus
\begin{align*}
  \sigma &= \left(\frac{\beta}{\alpha}\right)^{1/4} \left(\sqrt{\omega + 2\sqrt{\alpha\beta}} + \sqrt{\omega  
     - 2\sqrt{\alpha\beta}}\right), \\
  \tau &= \left(\frac{\alpha}{\beta}\right)^{1/4} \left(\sqrt{\omega + 2\sqrt{\alpha\beta}} - \sqrt{\omega - 
     2\sqrt{\alpha\beta}}\right), \\
  \varphi &= \left(\frac{\alpha}{\beta}\right)^{1/4} \left(\sqrt{\omega + 2\sqrt{\alpha\beta}} + \sqrt{\omega  
     - 2\sqrt{\alpha\beta}}\right), \\
  \chi &= \left(\frac{\beta}{\alpha}\right)^{1/4} \left(\sqrt{\omega + 2\sqrt{\alpha\beta}} - \sqrt{\omega - 
     2\sqrt{\alpha\beta}}\right), 
\end{align*}
yielding
\begin{equation*}
\begin{split}
  \q &= \frac{1}{\sqrt{\omega}} \biggl[\left(\gamma_+ \sqrt{\omega + 2\sqrt{\alpha\beta}} - \gamma_- 
     \sqrt{\omega - 2\sqrt{\alpha\beta}}\right) \omega x \\
  & \quad + {\rm i} \left(\gamma_- \sqrt{\omega + 2
     \sqrt{\alpha\beta}} - \gamma_+ \sqrt{\omega - 2\sqrt{\alpha\beta}}\right) p \biggr], \\
  \q^{\sharp} &= \frac{1}{\sqrt{\omega}} \biggl[\left(\gamma_+ \sqrt{\omega + 2\sqrt{\alpha\beta}} +
     \gamma_- \sqrt{\omega - 2\sqrt{\alpha\beta}}\right) \omega x \\
  & \quad + {\rm i} \left(\gamma_- \sqrt{\omega + 2 \sqrt{\alpha\beta}} + \gamma_+ \sqrt{\omega - 
     2\sqrt{\alpha\beta}}\right) p \biggr],     
\end{split}
\end{equation*}
with
\begin{equation*}
  \gamma_{\pm} = \frac{1}{2} \biggl[\left(\frac{\alpha}{\beta}\right)^{1/4} \pm \left(\frac{\beta}{\alpha}
  \right)^{1/4}\biggr]; 
\end{equation*}

\noindent (ii) for $z=1$, $\epsilon = - (\alpha - \beta)/[2(\omega - \alpha - \beta)]$ and thus
\begin{align*}
  \sigma &= \frac{1}{\sqrt{\omega-\alpha-\beta}} (\Omega + \omega - 2\beta), & \tau &= \frac{1}
     {\sqrt{\omega-\alpha-\beta}} (\Omega - \omega + 2\alpha), \\
  \varphi &= \frac{1}{\sqrt{\omega-\alpha-\beta}} (\Omega + \omega - 2\alpha), & \chi &= \frac{1}
     {\sqrt{\omega-\alpha-\beta}} (\Omega - \omega + 2\beta),
\end{align*}
yielding
\begin{align*}
  \q &= \frac{1}{\sqrt{\omega}} \left(\frac{\Omega+\alpha-\beta}{\sqrt{\omega-\alpha-\beta}}\, \omega x
     - {\rm i} \sqrt{\omega-\alpha-\beta}\, p\right), \\
  \q^{\sharp} &= \frac{1}{\sqrt{\omega}} \left(\frac{\Omega-\alpha+\beta}{\sqrt{\omega-\alpha-\beta}}\, 
     \omega x + {\rm i} \sqrt{\omega-\alpha-\beta}\, p\right);
\end{align*}

\noindent (ii) for $z=-1$, $\epsilon = (\alpha - \beta)/[2(\omega + \alpha + \beta)]$ and thus
\begin{align*}
  \sigma &= \frac{1}{\sqrt{\omega+\alpha+\beta}} (\Omega + \omega + 2\beta), & \tau &= - \frac{1}
     {\sqrt{\omega+\alpha+\beta}} (\Omega - \omega - 2\alpha), \\
  \varphi &= \frac{1}{\sqrt{\omega+\alpha+\beta}} (\Omega + \omega + 2\alpha), & \chi &= - \frac{1}
     {\sqrt{\omega+\alpha+\beta}} (\Omega - \omega - 2\beta),
\end{align*}
yielding
\begin{align*}
  \q &= \frac{1}{\sqrt{\omega}} \left(\sqrt{\omega+\alpha+\beta}\, \omega x - {\rm i} 
     \frac{\Omega-\alpha+\beta}{\sqrt{\omega+\alpha+\beta}}\, p\right), \\
  \q^{\sharp} &= \frac{1}{\sqrt{\omega}} \left(\sqrt{\omega+\alpha+\beta}\, \omega x + {\rm i} 
     \frac{\Omega+\alpha-\beta}{\sqrt{\omega+\alpha+\beta}}\, p\right).
\end{align*}
\par
%
%
\section{Generalized models}

Since the operators $V_{\pm}$, $W_{\pm}$, defined in (\ref{eq:V-W}), are the odd generators of an ${\rm su}(1,1/1) \sim {\rm osp}(2/2, \R)$ Lie superalgebra \cite{bars}, whose even generators are $K_0$, $K_+$, $K_-$, defined in (\ref{eq:swanson-K}), and
\begin{equation*}
  Y = \tfrac{1}{2} \left(\bp b - \tfrac{1}{2}\right),
\end{equation*}
we see that the su(1,1) Lie algebra characterizing the non-Hermitian oscillator (\ref{eq:swanson}) has been naturally extended to such a superalgebra when going to the quasi-Hermitian SUSYQM description (\ref{eq:pseudo-SUSY}). The quasi-Hermitian supersymmetric Hamiltonian (\ref{eq:swanson-SUSY}) can indeed be written as a linear combination
\begin{equation}
  H_{\rm S} = 2\omega K_0 + 2\alpha K_- + 2\beta K_+ + 2\Omega Y  \label{eq:swanson-SUSY-bis}
\end{equation}
of the superalgebra even generators, while $\q$ and $\q^{\sharp}$, as given in (\ref{eq:pseudo-Q-bis}), are expressed in terms of the odd generators.\par
%
%
{}For completeness' sake, let us recall that the nonvanishing commutation or anticommutation relations of $K_0$, $K_{\pm}$, $Y$, $V_{\pm}$ and $W_{\pm}$ \cite{scheunert} are given by (\ref{eq:com}) and
\begin{equation}
\begin{array}{ll}
  [K_0, V_{\pm}] = \pm \tfrac{1}{2} V_{\pm}, &\qquad [K_0, W_{\pm}] = \pm \tfrac{1}{2} W_{\pm}, 
  \\[0.3cm]
  [K_{\pm}, V_{\mp}] = \mp V_{\pm}, &\qquad [K_{\pm}, W_{\mp}] = \mp W_{\pm}, \\[0.3cm]
  [Y, V_{\pm}] = \tfrac{1}{2} V_{\pm}, &\qquad [Y, W_{\pm}] = - \tfrac{1}{2} W_{\pm}, \\[0.3cm] 
  \{V_{\pm}, W_{\pm}\} = K_{\pm}, &\qquad \{V_{\pm}, W_{\mp}\} = K_0 \mp Y,      
\end{array}  \label{eq:com-bis}
\end{equation}
while their Hermiticity properties are provided by (\ref{eq:Hermite}) and
\begin{equation}
  Y^{\dagger} = Y, \qquad V_{\pm}^{\dagger} = W_{\mp}.  \label{eq:Hermite-bis}
\end{equation}
\par
%
%
We are now in a position to generalize the results of section 2 to all ($\Z_2$-graded) Hamiltonians that can be written in the form (\ref{eq:swanson-SUSY-bis}), where the realization of the ${\rm su}(1,1/1)$ even generators $K_0$, $K_+$, $K_-$ and $Y$ may be arbitrary. From \cite{cq07}, it follows that such Hamiltonians are quasi-Hermitian with respect to the (trivially extended) positive-definite metric operator $\zeta_+ = \rho^2$ with $\rho$ given in (\ref{eq:rho}). We plan to show that there exist supercharges $\q$ and $\q^{\sharp}$ satisfying equation (\ref{eq:pseudo-SUSY}) with them and related to the same $\zeta_+$. Furthermore, $\q$ and $\q^{\sharp}$ can be expressed in the form (\ref{eq:pseudo-Q-bis}), where $V_{\pm}$ and $W_{\pm}$ are the superalgebra odd generators in the corresponding realization, while $\sigma$, $\tau$, $\varphi$, $\chi$ are given by equation (\ref{eq:sigma}). To prove this assertion, we may only use the superalgebra defining relations (\ref{eq:com}), (\ref{eq:Hermite}), (\ref{eq:com-bis}) and (\ref{eq:Hermite-bis}).\par
%
%
To start with, it is obvious that the first part of equation (\ref{eq:pseudo-SUSY}) is trivially satisfied. On inserting equations (\ref{eq:pseudo-Q-bis}) and (\ref{eq:swanson-SUSY-bis}) in the second part of the same and using equation (\ref{eq:com-bis}), we get the following four conditions on the parameters $\sigma$, $\tau$, $\varphi$, $\chi$,
\begin{equation*}
  \sigma \varphi + \tau \chi = 4\omega, \qquad \sigma \varphi - \tau \chi = 4\Omega, \qquad \sigma \chi 
  = 4\beta, \qquad \tau \varphi = 4\alpha.
\end{equation*}
It is then straightforward to check that these restrictions are fulfilled by the expressions given in (\ref{eq:sigma}).\par
%
%
It remains to show that $\q^{\sharp} = \zeta_+^{-1} \q^{\dagger} \zeta_+$, which due to (\ref{eq:Hermite-bis}) amounts to the condition
\begin{equation}
  \rho (\varphi V_- + \chi V_+) \rho^{-1} = \rho^{-1} (\sigma V_- + \tau V_+) \rho  \label{eq:condition}
\end{equation}
for real parameters $\sigma$, $\tau$, $\varphi$, $\chi$. In \cite{cq07}, we proved that $\rho$, defined in (\ref{eq:rho-exp}), can be factorized in either of the following two forms
\begin{equation*}
  \rho = \exp(p K_+) \exp(q K_0) \exp (p K_-) = \exp(p' K_-) \exp(q' K_0) \exp(p' K_+),  
\end{equation*}
where\footnote{In these equations, we correct a misprint in the corresponding result of \cite{cq07} ($e^{-q'/2}$ is replaced by $e^{q'/2}$).}
\begin{equation}
\begin{array}{ll}
  e^{-q/2} = \cosh \theta - \epsilon \frac{\ds \sinh \theta}{\ds \theta}, &\qquad p = \frac{\ds z \epsilon
    \sinh \theta/\theta}{\ds \cosh \theta - \epsilon \sinh \theta/\theta}, \\[0.3cm]
  e^{q'/2} = \cosh \theta + \epsilon \frac{\ds \sinh \theta}{\ds \theta}, & \qquad p' = \frac{\ds z \epsilon
    \sinh \theta/\theta}{\ds \cosh \theta + \epsilon \sinh \theta/\theta}.
\end{array}  \label{eq:p-q}
\end{equation}
On combining these expressions with the Baker-Campbell-Hausdorff formula and equation (\ref{eq:com-bis}), it is then straightforward to prove the relations
\begin{equation}
\begin{array}{ll}
  \rho V_+ \rho^{-1} = e^{q'/2} (V_+ + p' V_-), &\qquad \rho V_- \rho^{-1} = e^{-q/2} (V_- - p V_+),    
     \\[0.3cm]
  \rho^{-1} V_+ \rho = e^{-q/2} (V_+ - p V_-), &\qquad \rho^{-1} V_- \rho = e^{q'/2} (V_- + p' V_+).
\end{array}  \label{eq:BCH}
\end{equation}
Finally, equations (\ref{eq:p-q}) and (\ref{eq:BCH}) allow us to easily check that condition (\ref{eq:condition}) is fulfilled by the parameters given in (\ref{eq:sigma}). This completes the extension of the results of section 2 to the generalized Hamiltonians (\ref{eq:swanson-SUSY-bis}) with  equivalent Hermitian Hamiltonians expressed as $h_{\rm S} = h + 2 \Omega Y$, where $h$ is given in (\ref{eq:h}).\par
%
%
Let us now review a few examples of generalized models based on (\ref{eq:swanson-SUSY-bis}). This amounts to considering some physically-relevant realizations of the ${\rm su}(1,1/1)$ generators.\par
%
%
{}For completeness' sake, let us mention some trivial extensions of the realization considered in section 2. The first one consists in replacing the fermionic creation and annihilation operators $\bp$, $b$ by $2 \times 2$ matrices $\sigma_+ = \bigl(\begin{smallmatrix}  0&1 \\ 0&0 \end{smallmatrix}\bigr)$ and $\sigma_- = \bigl(\begin{smallmatrix}  0&0 \\ 1&0 \end{smallmatrix}\bigr)$, respectively. In the second one, instead of a single pair of bosonic and fermionic operators $\ap$, $a$, $\bp$, $b$, one uses $n$ commuting pairs $\ap_i$, $a_i$, $\bp_i$, $b_i$, $i=1$, 2,~\ldots, $n$, thus yielding \cite{bars}
\begin{equation}
  K_0 = \frac{1}{2} \biggl(\sum_i \ap_i a_i + \frac{n}{2}\biggr), \qquad K_+ = \frac{1}{2} \sum_i 
     a^{\dagger2}_i, \qquad K_- = \frac{1}{2} \sum_i a_i^2,  \label{eq:ndim}
\end{equation}
\begin{equation*}
  Y = \frac{1}{2} \biggl(\sum_i \bp_i b_i - \frac{n}{2}\biggr),
\end{equation*}
\begin{equation*}
  V_+ = \frac{1}{\sqrt{2}} \sum_i \ap_i \bp_i, \qquad V_- = \frac{1}{\sqrt{2}} \sum_i a_i \bp_i, \qquad 
     W_+ = \frac{1}{\sqrt{2}} \sum_i \ap_i b_i, \qquad W_- = \frac{1}{\sqrt{2}} \sum_i a_i b_i.
\end{equation*}
The resulting pseudo-Hermitian supersymmetric Hamiltonian (\ref{eq:swanson-SUSY-bis}) is then a non-Hermitian $n$-dimensional boson-fermion oscillator.\par
%
%
A somewhat more subtle realization uses spin operators and provides a non-Hermitian generalization of the three-dimensional harmonic oscillator with a constant spin-orbit coupling $\frac{1}{2} (\pb^2 + \omega^2 \xb^2) \pm \omega \left(\sigmab \cdot \llb + \frac{3}{2}\right)$, first studied by Ui and Takeda \cite{ui} and whose `accidental' degeneracies were explained by Balantekin \cite{balantekin}. In this case, the realization of the $\mbox{\rm su}(1,1/1)$ generators reads
\begin{gather*}
  Y = \frac{1}{2} \left(\sum_i \sigma_i L_i + \frac{3}{2}\right) 
     \begin{pmatrix} 1 & 0 \\
     0 & -1\end{pmatrix}, \quad
  V_+ = \frac{1}{\sqrt{2}}
     \begin{pmatrix} 0 & \sum_i \sigma_i \ap_i \\
     0 & 0\end{pmatrix}, \quad 
  V_- = \frac{1}{\sqrt{2}}
     \begin{pmatrix} 0 & \sum_i \sigma_i a_i \\
     0 & 0\end{pmatrix}, 
  \\
  W_+ = \frac{1}{\sqrt{2}}
     \begin{pmatrix} 0 & 0 \\
     \sum_i \sigma_i \ap_i & 0 \end{pmatrix}, \quad
  W_- = \frac{1}{\sqrt{2}}
     \begin{pmatrix} 0 & 0 \\
     \sum_i \sigma_i a_i & 0 \end{pmatrix} 
\end{gather*}
with $K_0$, $K_+$, $K_-$ as in (\ref{eq:ndim}) and $i$ running over 1, 2, 3.\par
%
%
Turning ourselves now to the rich realm of generalized Calogero models \cite{calogero}, which were already signalled in \cite{cq07} as a possible domain for useful applications of our su(1,1) approach to \cite{musumbu}, we may consider ${\cal N} = 2$ superconformal extensions of such models \cite{freedman, gala}. In the case considered by Freedman and Mende \cite{freedman}, for instance, the ${\rm su}(1,1/1)$ generators are given by
\begin{align*}
  K_0 &= \frac{1}{2\omega} \Biggl( - \frac{1}{2} \sum_i \nabla_i^2 + \frac{1}{2} g^2 \sum_{\substack{
     i, j \\ i \ne j}} \frac{1}{(x_i - x_j)^2} + \frac{1}{2} \omega^2 \sum_i x_i^2 + \frac{1}{2} g
     \sum_{\substack{i, j \\ i \ne j}} \frac{1}{(x_i - x_j)^2} [\bp_i, b_j - b_i]\Biggr), \\
  \begin{split}
  K_{\pm} &= \frac{1}{2\omega} \Biggl[ \frac{1}{2} \sum_i \nabla_i^2 - \frac{1}{2} g^2 \sum_{\substack
     {i, j \\ i \ne j}} \frac{1}{(x_i - x_j)^2} + \frac{1}{2} \omega^2 \sum_i x_i^2 \mp \omega \biggl(\sum_i
     x_i \nabla_i + \frac{n}{2}\biggr) \\
  &\quad - \frac{1}{2} g \sum_{\substack{i, j \\ i \ne j}} \frac{1}{(x_i - x_j)^2} [\bp_i, b_j - b_i]\Biggr],   
  \end{split} \\
  Y &= \frac{1}{4} \biggl(\sum_i [\bp_i, b_i] + g n(n-1)\biggr), \\
  V_{\pm} &= \frac{1}{2\sqrt{\omega}} \sum_i \biggl(\mp \nabla_i + \omega x_i \mp g \sum_{j \ne i}
     \frac{1}{x_i - x_j}\biggr) \bp_i, \\ 
  W_{\pm} &= \frac{1}{2\sqrt{\omega}} \sum_i \biggl(\mp \nabla_i + \omega x_i \pm g \sum_{j \ne i}
     \frac{1}{x_i - x_j}\biggr) b_i,  
\end{align*}
where $x_i$, $i=1$, 2,~\ldots, $n$, denote the coordinates of $n$ particles on a line and $\bp_i$, $b_i$, $i=1$, 2,~\ldots, $n$, are $n$ independent pairs of fermionic creation and annihilation operators. The corresponding Hermitian supersymmetric Hamiltonian reads $2 \omega (K_0 + Y)$. The analysis presented in this section extends this operator to non-Hermitian Hamiltonians (\ref{eq:swanson-SUSY-bis}) with Hermitian counterparts.\par
%
%
\section{Final remarks}

In the present paper, by using an ${\rm su}(1,1/1) \sim {\rm osp}(2/2, \R)$ approach, we have further extended the set of non-Hermitian Hamiltonians to which the family of positive-definite metric operators constructed in \cite{musumbu} for a non-Hermitian oscillator can be applied. This has enlarged the class of non-Hermitian Hamiltonians related to Hermitian ones by similarity transformations and therefore endowed with solid physical foundations.\par
%
%
In particular, we have proposed non-Hermitian generalizations of the Ui and Takeda model and of the Freedman and Mende ${\cal N} = 2$ superconformal extension of the Calogero model. These are but some simple examples of relevance of our new proposal, which will undoubtedly prove rich in applications to many fields.\par
%
%
As a final point, it is worth commenting on some specific physical aspects of the non-Hermitian Hamiltonians considered here, which may shed some light on the algebraic construction just carried out. In the literature, one can find other non-Hermitian Hamiltonians with a harmonic oscillator spectrum, for which the non-Hermiticity lies in a complex-valued potential which is purely local. Such is the case, for instance, for some \pt-symmetric (or non-\pt-symmetric) Hamiltonians constructed  by using a SUSYQM or Darboux technique \cite{cannata98}. In contrast, for the non-Hermitian oscillator Hamiltonian (\ref{eq:swanson}), the non-Hermiticity is due to a momentum-dependent linear interaction $\frac{1}{2} {\rm i} (\alpha-\beta) (xp+px)$. The gauge-type transformation performed by $\rho$, defined in (\ref{eq:rho-exp}),\footnote{It should be noted that $\rho$ itself depends on $p$ for $z \ne 1$, since equation (\ref{eq:rho-exp}) can be rewritten as $\rho = \exp \{\epsilon [(1-z) p^2 + (1+z) \omega^2 x^2]/(2\omega)\}$.} eliminates such a non-locality while restoring Hermiticity. So we may say that for such a Hamiltonian cure for non-Hermiticity is equivalent to cure for non-locality. Of course, the underlying su(1,1) symmetry is largely instrumental in this matter. Such remarks remain true for the supersymmetric extension (\ref{eq:swanson-SUSY}) (as well as for the generalizations (\ref{eq:swanson-SUSY-bis}) of the latter) because it only differs from (\ref{eq:swanson}) by a Hermitian term $2\Omega Y$.\par  
%
%
\section*{Acknowlegments}

The author is grateful to A Galajinsky for bringing reference \cite{gala} to her attention and to an anonymous referee for some interesting comments and suggestions.\par
%
%

\end{document}